\documentclass[aps]{revtex4}
\usepackage{color}
\usepackage{dcolumn}
\usepackage{bm}
\usepackage{float}
\usepackage[latin1]{inputenc}
\usepackage[spanish,english]{babel}
\usepackage{amsfonts}
\usepackage{graphicx}
\usepackage{amsmath,amssymb}
\usepackage{hyperref}

\begin{document}
	
	\title{\textbf Isotropic and anisotropic neutron star structure in 4D Einstein-Gauss-Bonet Gravity }
	\author{ Gholam Hossein Bordbar\footnote{%
			email address: ghbordbar@shirazu.ac.ir}, Mohammad Mazhari and Ahmad Poostforush}
	\affiliation{Physics Department and Biruni Observatory, Shiraz University, Shiraz 71454, Iran}

	
\begin{abstract}

With regards to the coupling constant and the strong magnetic field of neutron stars, we have studied these stars in the 4D Einstein-Gauss-Bonnet (4D EGB) gravity model in order to grasp a better understanding of these objects. In this paper, we have shown that the neutron stars properties are considerably affected by the coupling constant and magnetic field. We have found that as a consequence of the strong magnetic field and the coupling constant, the maximum mass and radius of a neutron star are increasing functions of the coupling constant, while Schwarzschild radius, compactness, surface gravitational redshift, and Kretschmann scalar are decreasing functions. Additionally, our study has shown that the physical properties of a magnetized neutron star not only are greatly influenced by the strong magnetic field, but also by the anisotropy. Moreover, we have shown that to obtain the hydrostatic equilibrium configuration of the magnetized material, both the local anisotropy effect and the anisotropy due to the magnetic field should be considered. Finally, we have found that in the anisotropic magnetized neutron stars, the maximum mass and radius do not always increase with increasing the internal magnetic field.

	
\end{abstract}

\maketitle

\section{Introduction}
A neutron star is an ideal astrophysical laboratory for testing condensed matter physics theories and making connections between nuclear physics, particle physics, and astrophysics \cite{Steiner,Lattimer}. This makes the neutron star as a valuable system to study its various properties. The purpose of present work is to evaluate the structural characteristics of isotropic and anisotropic magnetized neutron stars containing pure neutron matter. Recent observations have revealed that the neutron stars have a mass $M\geq 2.5M_{\odot }$ \cite{abbott17,abbott19}.

Owing to the existence of a strong magnetic field, there is anisotropy in the magnetized neutron star pressure; therefore, the  pressure is different in the radial ($P_{r}$) direction from the tangential ($P_{t}$) direction.
According to available evidence \cite{Yuan,Tatsumi,Ferrer}, it is difficult to obtain the strength of the magnetic field inside the magnetized neutron stars, prompting researchers to develop theoretical models that can be used to analyze the influence of strong magnetic fields on the physical parameters of magnetized neutron stars.
Magnetized neutron stars are classified into two groups based on their surface magnetic field. Pulsars and magnetars have different magnetic fields on their surface.
Pulsars are found to have surface magnetic fields of approximately $10^{13}G$ to $10^{15}G$ \cite{Duncan}, and magnetars are found to have surface magnetic fields of approximately $10^{16}G$ to $10^{17}G$ \cite{Cenko}.

Einstein-Gauss-Bonnet gravity is a natural generalization of general relativity. As the most general torsion-free theory of gravity, this leads to stable second-order equations of motion in higher dimensions.
A version of this theory was proposed by Lanczos \cite{Lanczos}, and later confirmed by David Lovelock \cite{Lovelock1,Lovelock2}.
Because this theory shares Einstein's principle of gravity, the Einstein-Gauss-Bonnet (EGB) gravity is more suitable than other higher-curvature gravity theories.
Einstein's and GB's terms are included in Lovelock's Lagrangian theory. As a result, the famous EGB gravity was proposed, which is the most straightforward nontrivial generalization of Einstein's gravity.
It has been shown that static wormhole solutions in the EGB theory of gravity with higher dimensions can be found \cite{Maeda,Jusufi}.

{A notable feature of EGB gravity in the 4-dimensional case is that the Euler-Gauss-Bonnet term becomes a topological invariant that contributes neither to equations of motion nor gravitational dynamics \cite{eslamjaf}.}
A recent paper by Glavan and Lin \cite{Glavan} proposes a modified general covariant theory of gravity in 4-spacetime dimensions ($D = 4$) that propagates only the massless graviton and bypasses Lovelock's theorem. The Einstein gravity theory with the cosmological constant is the only gravity theory that meets several conditions; metricity, diffeomorphism invariance, and second-order equations of motion. Essentially, this theory is formulated in higher dimensions, $D > 4$, and has Einstein-Hilbert terms with cosmological constants, as well as GB coupling rescaled as $\alpha/(D-4)$. 4D theory is defined as $D \rightarrow 4$.
{As a result, the GB invariance gives rise to the nontrivial contributions in  gravitational dynamics, at the same time preserving the graviton degrees of freedom and retaining the   freedom from Ostrogradsky instability \cite{eslamjaf}. This new theory is called 4D EGB gravity (see Ref. \cite{Glavan} for more details).}

{In other gravities such as $f(\mathcal{R})$ gravity, it is specifically discussed the mass-radius diagram for static neutron star models obtained by the numerical solution of modified Tolman-Oppenheimer-Volkoff equations \cite{Capozziello}. This study explores the impact of different Lagrangians, including quadratic and cubic corrections, on the mass-radius relation. In addition, it addresses the constraints imposed by the presence of an extra degree of freedom and the stiffness of the equation of state, providing valuable insights into the mass-radius relation for neutron stars in the context of $f(\mathcal{R})$ gravity. There is also the possibility of discussing strong gravitational fields in local objects in $f(\mathcal{R})$ gravity due to its higher-order curvature terms. Neutron stars are frequently studied within this framework \cite{Capozziello}. An inflationary phase of the universe (the earliest phases of the Universe) can also be unified by $f(\mathcal{R})$ theory \cite{Yousaf,Bhatti}. In addition, for $f(\mathcal{R,T})$ gravity, the widest parameters and masses are investigated to ensure the stability and stability of static wormhole models \cite{Bhatti2}.}

Neutron stars are denser and more compact astronomical bodies than ordinary stars, and are strongly dependent on their equation of state (i.e., the relationship between the density and pressure).
It is unknown what a neutron star comprises, and how it is structured internally. Because of this lack of knowledge, physicists have to resort to theoretical assumptions on different equations of state to determine the characteristics of neutron stars. In view of this, several models have been proposed by other research groups for the equations of state of these objects. Despite this, few of them are stable and realistic for describing the matter inside these compact bodies. The problem of compact stars made of anisotropic material has been studied in Refs. \cite{DEB,Doneva,4Tangphati,5Tangphati}.

Our previous researchs examined neutron star structure properties in the absence and presence of a magnetic field using an isotropic neutron star model. According to the contemporary equation of state, using the equation of state based on the current nuclear potential AV18 \cite{bordbar1998}, we calculated the neutron star structure \cite{Bordbar1}, and a comparison was made between the structure of cold and hot neutron stars with quark cores and those without a quark core \cite{Bordbar2, Bordbar7}. In addition, we examined the effects of the cosmological constant $\Lambda$ and other gravity theories such as (3 + 1)-dimensional rainbow gravity and spin-2 massive gravitons on the properties of neutron stars \cite%
{Bordbar3,Bordbar4,Bordbar5,Bordbar6}.
Afterwards, we evaluated the properties of cold neutron stars with a quark core \cite{Bordbar7}.
The properties of spin polarized neutron matter
in the presence of strong magnetic fields at zero \cite{Bordbar8,Bordbar9,Bordbar10}, and finite temperatures \cite{Bordbar11} have also been studied.

It has been attempted many times to conceive of spherically symmetric fluid spheres with anisotropic matter as suitable  internal models for these stars. The Bower and Liang model \cite{BowerLiang}, which considers radial pressure as part of the anisotropic function, strengthened the idea of anisotropic materials.
Despite this, Ruderman \cite{Ruderman} observed that nuclear matter becomes anisotropic at densities significantly greater than $10^{15}g/cm^3$.
Moreover, anisotropic pressure affects the mass-radius relation in compact stars \cite{Horvat,Karami,Sedaghat}.
A strong magnetic field appears to alter the maximum mass of a compact stellar object, but researchers \cite{Bandyopadhyay1,Bandyopadhyay2,Casali,Broderick,Kayanikhoo} remain split on whether it increases or decreases the maximum mass, a difficult open question.

In the present study, we evaluate the structural properties of an anisotropic neutron star containing pure neutron matter and also an isotropic neutron star that is not magnetized. We use the lowest-order constrained variational (LOCV) method to calculate the equation of state of neutron matter for non-magnetized and magnetized neutron stars in the absence and presence of a magnetic field. As a final step, we solve the equations in 4D EGB gravity to determine the maximum mass, radius, and other properties of the system using this equation of state.

\section{Equation of state of isotropic non- magnetized and anisotropic magnetized neutron star}
The equation of state is necessary to investigate the properties of isotropic non-magnetic neutron stars and anisotropic magnetic neutron stars, allowing us to calculate the structural properties of the star using this equation.
In this section, we calculate the equation of state for isotropic non-magnetized and anisotropic magnetized neutron stars.
Here, at first we need to have the total energy of the system.
In order to calculate the energy, we use the  LOCV method.
In the LOCV formalism, we consider the wave function of the interacting system as follows,
\begin{equation}
	\psi =F \phi.
\end{equation}
Here, $\phi$ is the wave function of $N$ non-interacting neutrons and $F$ is the $N$-body correlation operator, which is considered as follows according to the Jastrow approximation,
\begin{equation}
	F =S\prod_{i>j}f(ij),
\end{equation}
where $S$ and $f(ij)$ are symmetry operator and two-body correlation function, respectively.
For calculating the Hamiltonian expectation value, we use the cluster expansion, and keep only the one-body and two-body terms,
\begin{equation}
	E([f])=\frac{\langle {\psi }|H|{\psi }\rangle }{N \langle {\psi }|{%
			\psi }\rangle }=E_{1}+E_{2}.
\end{equation}
Here, $H$ denotes the Hamiltonian of system, and $E_{1}$ is the kinetic energy of Fermi gas which is calculated as follows,
\begin{equation}
	E_{1}=\sum_{i=+,-}\frac{3}{5}\frac{\hbar ^{2}k_{F}^{{(i)}^{2}}}{2m}\frac{%
		\rho ^{(i)}}{\rho },
\end{equation}%
where $k_{F}^{(i)}={(6\pi ^{2}\rho ^{(i)})}^{(\frac{1}{3})}$ is the Fermi
momentum of a neutron with spin projection $i$. The two-body energy $E_{2}$ is also obtained as follows,
\begin{equation}
	E_{2}=\frac{1}{2N}\sum_{ij}\langle {ij}|\nu {(12)}|{ij-ji}\rangle ,
\end{equation}%
where
\begin{equation}
	\nu {(12)}=-\frac{\hbar ^{2}}{2m}[f(12),[\bigtriangledown
	_{12}^{2},f(12)]]+f(12)V(12)f(12).
	\label{nu}
\end{equation}
In the above equation, $V(12)$ is the two-body interaction potential. In the Ref. \cite{bordbar1998}, the detail of calculations for nuclear matter equation of state has been given.

In the presence of a strong magnetic field, the contribution of magnetic energy for the magnetized neutron matter is added to the total energy as follows,
\begin{equation}
	E(\rho ,B)=E_{1}+E_{2}-\mu _{n}B\delta.
	\label{r}
\end{equation}
In the above equation, $\mu _{n}$, $B$ and $\delta$ are the neutron magnetic momentum, magnetic field, and spin polarization parameter, respectively.
To obtain the equation of state of the neutron star matter, at each magnetic field, we use the following equation,
\begin{equation}
	P(\rho ,B)=\rho ^{2}\biggl(\frac{\partial E(\rho ,B)}{\partial \rho }\biggr)%
	_{B}.  \label{press}
\end{equation}%
In a strong magnetic field, the pressure of neutron star differs in the radial and tangential directions, indicating that the neutron star undergoes the anisotropy. From the following equation, we can calculate the tangential ($P_{t}$) and radial ($P_{r}$) pressure components, respectively \cite{Mallick},
\begin{equation}
	P_{t}=\rho ^{2}\biggl(\frac{\partial E(\rho ,B)}{\partial \rho }\biggr)_{B}+%
	\frac{B^{2}}{8\pi },
\end{equation}%
\begin{equation}
	P_{r}=\rho ^{2}\biggl(\frac{\partial E(\rho ,B)}{\partial \rho }\biggr)_{B}-%
	\frac{B^{2}}{8\pi }.
	\label{Pr}
\end{equation}

In order to examine the effect of magnetic field, we here consider a Gaussian function of the density as follows \cite{Bandyopadhyay1},
\begin{equation}
	B(\rho)=B_{surf}+B_{0}[1-\exp (-{\beta (\frac{\rho }{\rho _{0}}%
		)^{\theta} )}],
\end{equation}%
where $\rho$ is the neutrons density, $\rho_{0}$ is the saturation density, $B_{surf}$ is the magnetic field on the star's surface, which is equal to $10^{15} G$, and $B_{0}$ is the magnetic field in the dense (central) region of the star.
There are two behaviors for the magnetic field based on the remaining parameters $\beta$ and $\theta$. For fast decay, $\beta=0.02$ and  $\theta=3.0 $ are considered, and $\beta=0.05 $ and  $\theta=2.0 $ are considered for slow decay.
In a fast decay (slow decay) of the magnetic field, the field drops rapidly (slowly) from the star center to its surface (see Ref. \cite{Casali} for more details).
Now, using the Gaussian magnetic field, as mentioned previously, we calculate the equation of state of the system.

In the absence of a magnetic field ($B=0$), the radial and tangential components of pressure are equal, so there is no anisotropy in the neutron star. In addition, the difference in the tangential and radial pressures is not significant in the case of weak magnetic fields.
\begin{figure}[h!]
	\begin{center}{\includegraphics[width=9.5cm]{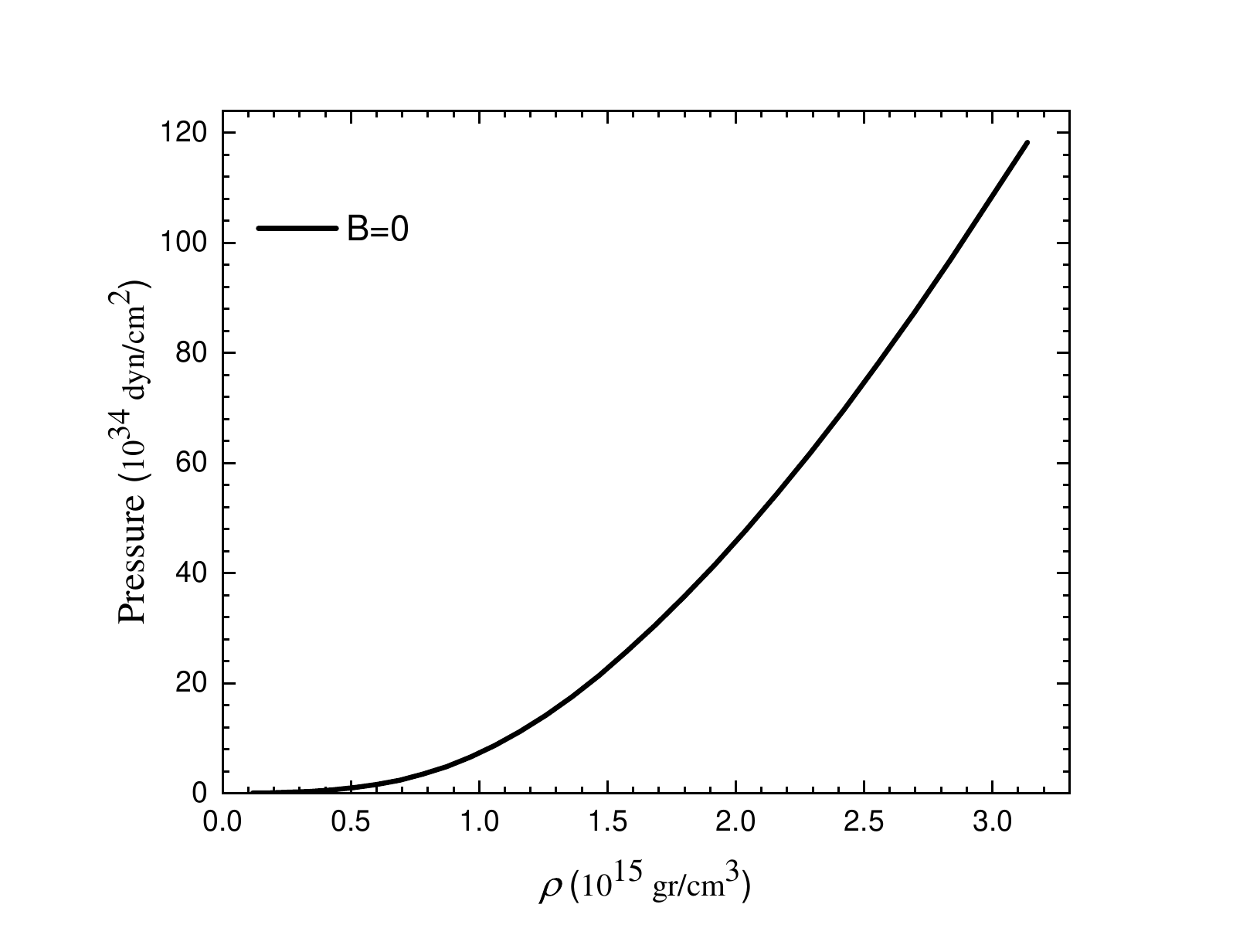}
			\includegraphics[width=9.5cm]{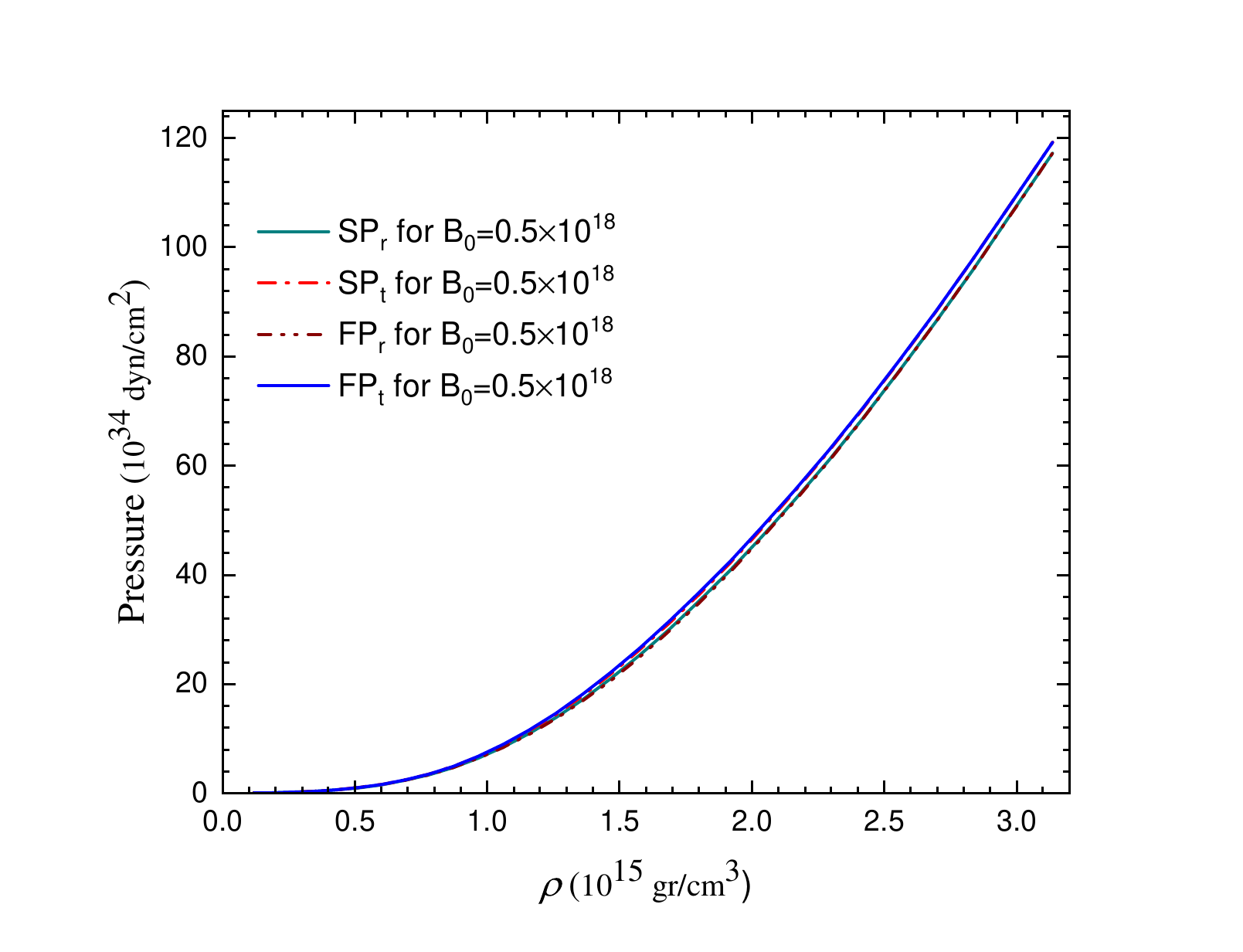}}
		\caption{{\protect\small The equation of state of non-magnetized neutron star
				for $B=0$ (up panel) and (down panel) the radial pressure and tangential
				pressure vs density of magnetized neutron star for two mode slow-decay (S) and fast-decay (F) with $B_{0}=0.5\times10^{18}G$.}}
		\label{EOS1}
	\end{center}
\end{figure}
\begin{figure}[h!]
	\begin{center}{\includegraphics[width=9.5cm]{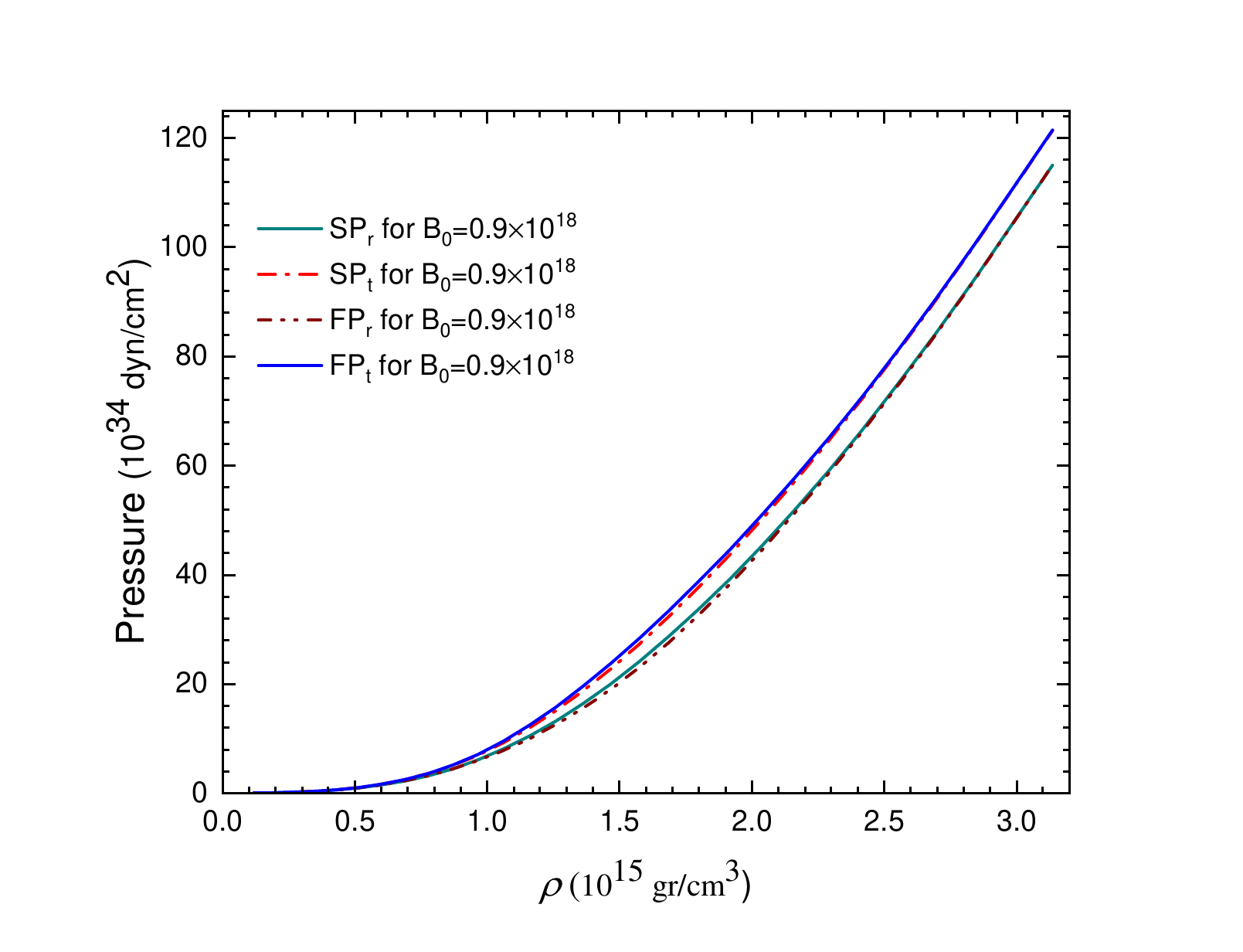}
			\includegraphics[width=9.5cm]{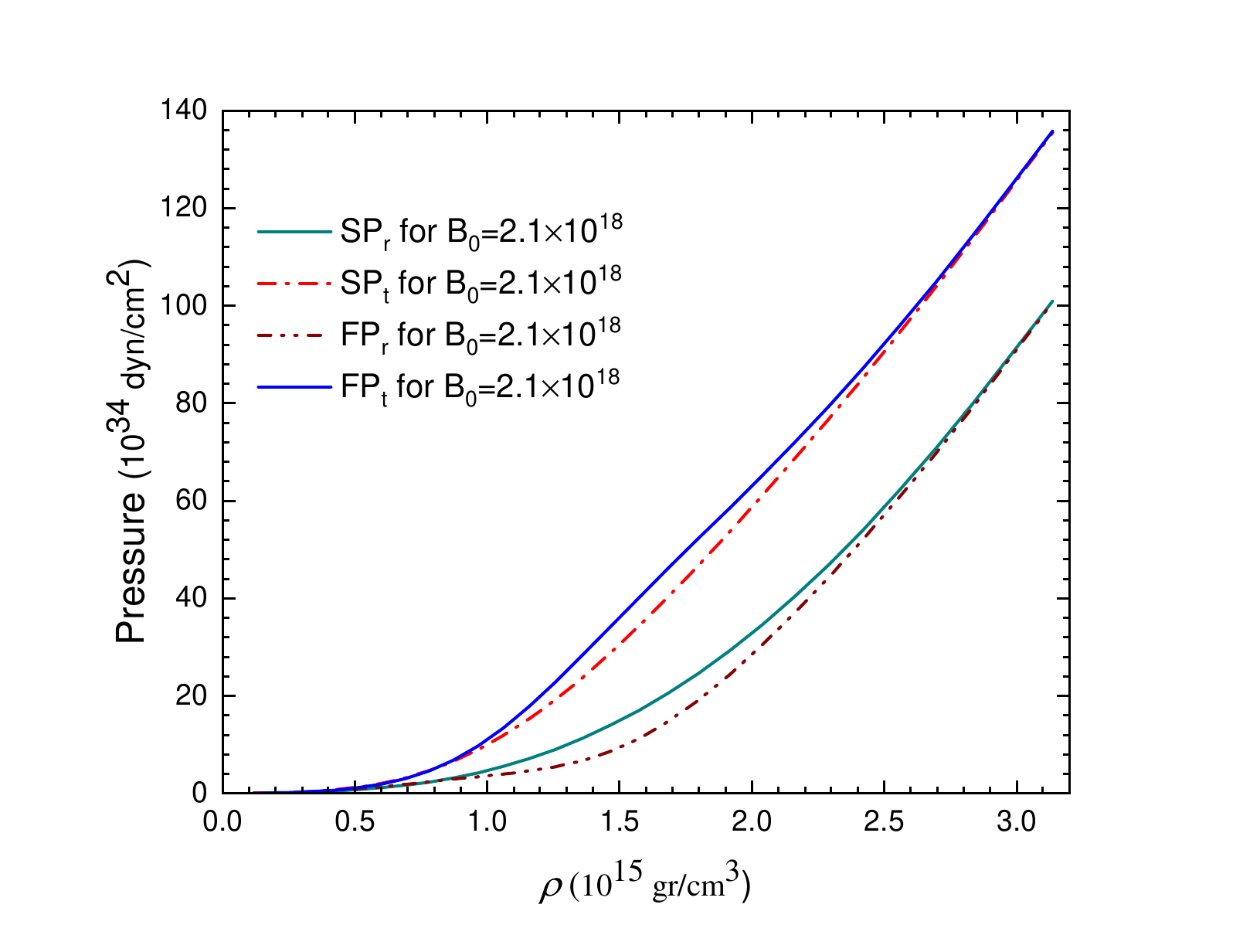}}
		\caption{{\protect\small The radial pressure and tangential pressure vs
				density of magnetized neutron star for $B=0.9\times10^{18}G$ (up panel) and
				for two mode slow-decay (S) and fast-decay (F) with $B_{0}=2.1\times10^{18}G$ (down panel). }}
		\label{EOS2}
	\end{center}
\end{figure}
Nevertheless, when the magnetic field increases, the anisotropy becomes clear under the pressure of the neutron star. There is a clear indication of this phenomena in Figs. \ref{EOS1} and \ref{EOS2}.
For different internal magnetic fields $B_{0}$ in two cases of the Gaussian magnetic field decay (fast decay and slow decay), we plot the equations of state in Figs. \ref{EOS1} and \ref{EOS2}.
It is clear from these graphs that as the internal magnetic field of the neutron star increases, the anisotropy in the magnetized neutron star increases too.
As the central density increases, there is a rising difference between the radial and tangential pressures. The pressure in the tangential component increases at a slower rate in slow decay than in fast decay, and the pressure in the radial component decreases at a slower rate in slow decay than in fast decay for intermediate values of the central density. To use the equations of the isotropic state and the anisotropic magnetized state (the radial pressure component of the slow decay is considered here), it is necessary to investigate the causality condition.
In a physically acceptable model, the speed of sound ($\nu$) should be less than the speed of light ($c$) to fulfill the causality conditions for the equation of state of neutron stars.
\begin{figure}[h!]
	\begin{center}{\includegraphics[width=9.5cm]{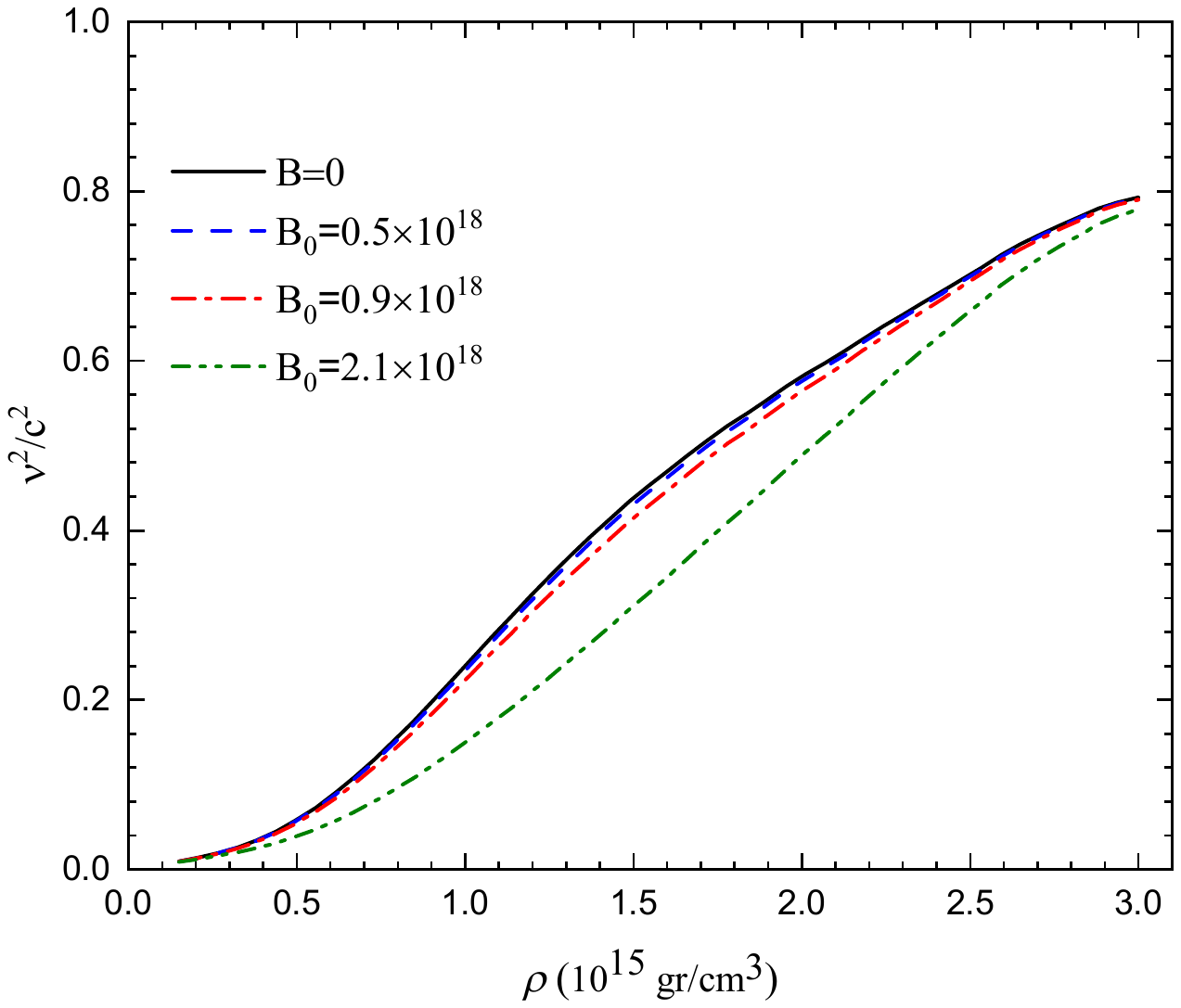}}
		\caption{{\protect\small Sound speed vs central density for different magnetic fields. }}
		\label{velocity}
	\end{center}
\end{figure}
Therefore, the condition of causality is given as follows,
\begin{equation}
	0\leqslant \frac{\nu ^{2}}{c^{2}}=\frac{1}{c^{2}}(\frac{dP_{r}}{d\rho }) \leqslant 1
\end{equation}
We present our results for the speed of sound in terms of the central density in Fig. \ref{velocity}. Accordingly, the magnetized neutron star equation of state also fulfills the causality condition for the magnetic fields mentioned above (see Fig. \ref{velocity} for details). In accordance with the results presented in Ref. \cite{Tews}, the equation of state of anisotropic magnetized neutron stars is relevant for studying their structural properties, such as gravitational mass and corresponding radius.
As discussed above, the equations of the state we are considering satisfy the causality conditions, therefore we can use them in the structure calculations.

It is well known that to compute the structural properties of a compact object, we must solve the  general relativistic hydrostatic equilibrium equations with the desired gravity.
In the next section, we discuss the basic formalism and modified hydrostatic equilibrium equations for 4D EGB gravity.

\section{The structure of isotropic non-magnetized neutron star in 4D EGB gravity}
In this section, we compute the structure of an isotropic non-magnetized neutron star in 4D EGB gravity. A brief overview of 4D EGB gravity is given in the following part,
 followed by a presentation of the modified Tolman-Oppenheimer-Volkoff (TOV) equations for the stellar structure of isotropic compact stars.

\subsection{MODIFIED TOV EQUATIONS IN 4D EGB
	GRAVITY}
To obtain the field equations for 4D EGB gravity, we consider the D-dimensional action as follows \cite{Glavan},
\begin{equation}
	S=\frac{c^4}{16\pi G}\int_{M} d^{D}x \sqrt{-g} \biggl(R+\frac{\alpha}{D-4}\mathcal{L}_{GB}\biggr)+S_{matter}.
	\label{S4}
\end{equation}
As shown in Eq. (\ref{S4}), we have rescaled the coupling constant $\alpha \rightarrow \alpha/(D-4)$. Therefore, the above theory is free of Ostrogradski instability, and a novel 4D EGB gravity can be defined in the limit $D \rightarrow 4$. In Eq. (\ref{S4}), $R$ is the Ricci scalar and $g$ is the determinant of the metric tensor $g_{\mu\nu}$. The GB coupling constant $\alpha$ has dimension of $[length]^2$, and $\mathcal{L}_{GB}$ is the GB term defined by
\begin{equation}
	\mathcal{L}_{GB} \equiv R^{\mu \nu \rho \sigma}R_{\mu \nu \rho \sigma}-4R^{\mu \nu}R_{\mu \nu}+R^{2}.
\end{equation}
In Eq. (\ref{S4}), $S_{matter}$ represents the distribution of matter lagrangian, which is independent of its derivatives, and depends only on the metric tensor components $g_{\mu\nu}$. The variation of Eq. (\ref{S4}) in terms of the metric tensor $g_{\mu\nu}$, leads to the following field equations,
\begin{equation}
	G_{\mu \nu}+\frac{\alpha}{D-4}H_{\mu \nu}=\frac{8\pi G}{c^4} T_{\mu \nu},
	\label{eqgb4}
\end{equation}
where the Einstein tensor $G_{\mu \nu}$ and Lanczos tensor $H_{\mu \nu}$ are defined by
\begin{equation}
	G_{\mu \nu} \equiv R_{\mu \nu}-\frac{1}{2} R g_{\mu\nu},
	\label{Gmunu}
\end{equation}
\begin{eqnarray}\nonumber
	H_{\mu \nu} &\equiv& 2 \biggl(RR_{\mu \nu}-2R_{\mu \sigma} R^{\sigma}_{\nu}-2R_{\mu \sigma \nu \rho}R^{\sigma \rho}-
R_{\mu \sigma \rho \delta}R^{\sigma \rho \delta}_{\nu} \biggr)\\
	&&-\frac{1}{2} g_{\mu \nu}\mathcal{L}_{GB}.
	\label{Hmunu}
\end{eqnarray}
The energy-momentum tensor is given by
\begin{equation}
	T_{\mu\nu}=-\frac{2}{\sqrt{-g}}\frac{\delta(\sqrt{-g}L_{m})}{\delta g^{\mu \nu}},
	\label{Tmuna}
\end{equation}
where $T_{\mu\nu}$ is the energy-momentum tensor of the matter field, $R$ is the Ricci scalar, $R_{\mu \nu}$ the Ricci tensor and $R_{\mu \sigma \nu \rho}$ being the Riemann tensors, respectively. When $\alpha=0$, it is easy to see that Eq. (\ref{eqgb4}) reduces to the conventional Einstein field equation.
The corresponding line element is expressed as follows in the form of a spherically symmetric $4$-dimensional metric describing the interior spacetime of non-rotating neutron stars,
\begin{equation}
	ds^{2}_{4}=-e^{2\Phi(r)}(cdt)^{2}+e^{2\lambda(r)}dr^{2}+r^{2}d\Omega^{2}_{2},
	\label{metric}
\end{equation}
where the metric functions $\Phi(r)$ and $\lambda(r)$ depend only on the radial coordinate $r$, and $d\Omega^{2}_{2}$ represents the metric on
the surface of $2$-sphere, namely,
\begin{eqnarray}\nonumber
	d\Omega^{2}_{2}&=&d\theta^{2}_{1}+sin^{2}\theta_{1} d\theta^{2}_{2}.
\end{eqnarray}
In addition, we assume that the matter source is described by an isotropic fluid whose energy-momentum tensor is given by
\begin{equation}
	T_{\mu \nu}=(\epsilon+P)u_{\mu}u_{\nu}+Pg_{\mu \nu},
\end{equation}
 where $u_{\mu}$ is the $4$-velocity of the fluid, $\epsilon=c^{2}\rho$ is the energy density of standard matter, $\rho$ is the mass density, and $P$ is the pressure. Here, $u^{\mu}$ is defined as $u^{\mu}=(e^{-\Phi(r)},0,0,0)$ with $u^{\mu}u_{\mu}=-1$.

According to Eq. (\ref{metric}), the effective 4D EGB theory can be obtained in the singular limit $D \rightarrow 4$ with  $\alpha \rightarrow \alpha/(D-4)$. By carefully performing the limit $D \rightarrow 4$, we obtain the following dimensionally reduced field equations \cite{Doneva},
\begin{eqnarray}
\frac{2}{r}\frac{d\lambda}{dr}=e^{2\lambda}\frac{\frac{8\pi G \rho}{c^2}-\frac{(1-e^{-2\lambda})}{r^2}(1-\alpha \frac{(1-e^{-2\lambda})}{r^2})}{(1+2\alpha\frac{(1-e^{-2\lambda})}{r^2})},
\end{eqnarray}%
\begin{eqnarray}
	\frac{2}{r}\frac{d\Phi}{dr}=e^{2\lambda}\frac{\frac{8\pi G P}{c^4}+\frac{(1-e^{-2\lambda})}{r^2}(1-\alpha \frac{(1-e^{-2\lambda})}{r^2})}{(1+2\alpha\frac{(1-e^{-2\lambda})}{r^2})}.
\end{eqnarray}%
In addition, the non-trivial radial component ($\nu=1$) of the conservation law of the isotropic energy-momentum tensor (continuity equation) obtained from the relation $\nabla_{\mu}T^{\mu}_{\hspace{2mm}\nu}=\partial_{\mu}T^{\mu}_{\hspace{2mm}\nu}+\Gamma^{\mu}_{\rho\mu}T^{\rho}_{\hspace{2mm}\nu}-\Gamma^{\rho}_{\mu\nu}T^{\mu}_{\hspace{2mm}\rho}=0$, is given by the following formula,
\begin{eqnarray}
	\frac{dP}{dr}=-(\rho c^2 +P )\frac{d\Phi}{dr}.
\end{eqnarray}%
	
To study the general structure of the solution, we replace the metric function ($e^{-2\lambda(r)}$) by $e^{-2\lambda(r)}=1-\frac{r^2}{2 \alpha G}\bigg(\sqrt{1+\frac{8\alpha G^2 m(r)}{c^2 r^3}}-1\bigg)$ (see Ref. \cite{Doneva}, for more details).
{There are also some authors who convenient adapt the asymptotic function of the metric function $e^{-2\lambda(r)}\approx1-\frac{2Gm(r)}{c^2r}$ for computing the compact object structure. This point has been fully discussed in Ref. \cite{Astashenok}.}
After some algebra, the generalized TOV equations in 4D EGB gravity for an isotropic neutron star with spherical symmetry in hydrostatic equilibrium are as follows,

\begin{eqnarray}\nonumber
	\frac{dP}{dr}&=&-\frac{r}{2}\frac{( \rho
		(r)c^{2}+ P(r) )}{1-\frac{r^2}{2 \alpha G}\bigg(\sqrt{1+\frac{8\alpha G^2 m(r)}{c^2 r^3}}-1\bigg)} \\&&\times\frac{\frac{8\pi GP(r)}{c^4}+\frac{1}{2\alpha G}\bigg(\sqrt{1+\frac{8\alpha G^2 m(r)}{c^2 r^3}}-1\bigg)\biggl(1-\frac{1}{2G}\bigg(\sqrt{1+\frac{8\alpha G^2 m(r)}{c^2 r^3}}-1\bigg)\biggr)}{1+\frac{1}{G}\bigg(\sqrt{1+\frac{8\alpha G^2 m(r)}{c^2 r^3}}-1\bigg)},
\end{eqnarray}%
\begin{eqnarray}\nonumber
	\frac{dm(r)}{dr}&=&-\frac{m(r)}{r}+\frac{c^2 r^2}{2G}\biggl(\frac{\frac{8\pi G \rho(r)}{c^2}-\frac{1}{2\alpha G}\bigg(\sqrt{1+\frac{8\alpha G^2 m(r)}{c^2 r^3}}-1\bigg)\biggl(1-\frac{1}{2G}\bigg(\sqrt{1+\frac{8\alpha G^2 m(r)}{c^2 r^3}}-1\bigg)\biggr)}{1+\frac{1}{G}\bigg(\sqrt{1+\frac{8\alpha G^2 m(r)}{c^2 r^3}}-1\bigg)}\biggr)\\&&+\frac{c^2 r^2}{2 \alpha G^2 }\bigg(\sqrt{1+\frac{8\alpha G^2 m(r)}{c^2 r^3}}-1\bigg), \label{tov2}
\end{eqnarray}%
where $\rho (r)$ is the energy density, $G$ is the gravitational constant, $\alpha$ is the  the coupling constant of the Gauss-Bonnet term and $m(r)$ is the gravitational mass inside the radius $r$. Thus, by taking $\alpha = 0$, we return to the TOV equations for a neutron star in Einstein gravity \cite{Oppenheimer}.

Here, to determine the structure of a neutron star, we use the equation of state for a nonmagnetized isotropic neutron star and generalized isotropic TOV equations with boundary conditions ($P(0)=P_{c}$ and $m(0)=0$). By integrating various coupling constant values from the center of the star $r=0$ to its surface $r=R$, where the pressure is zero, we obtain the maximum mass and the corresponding radius of the star. The following section presents our numerical results and discusses the structural properties of neutron stars in 4D EGB gravity.
\begin{figure}[h!]
	\begin{center}{\includegraphics[width=9.5cm]{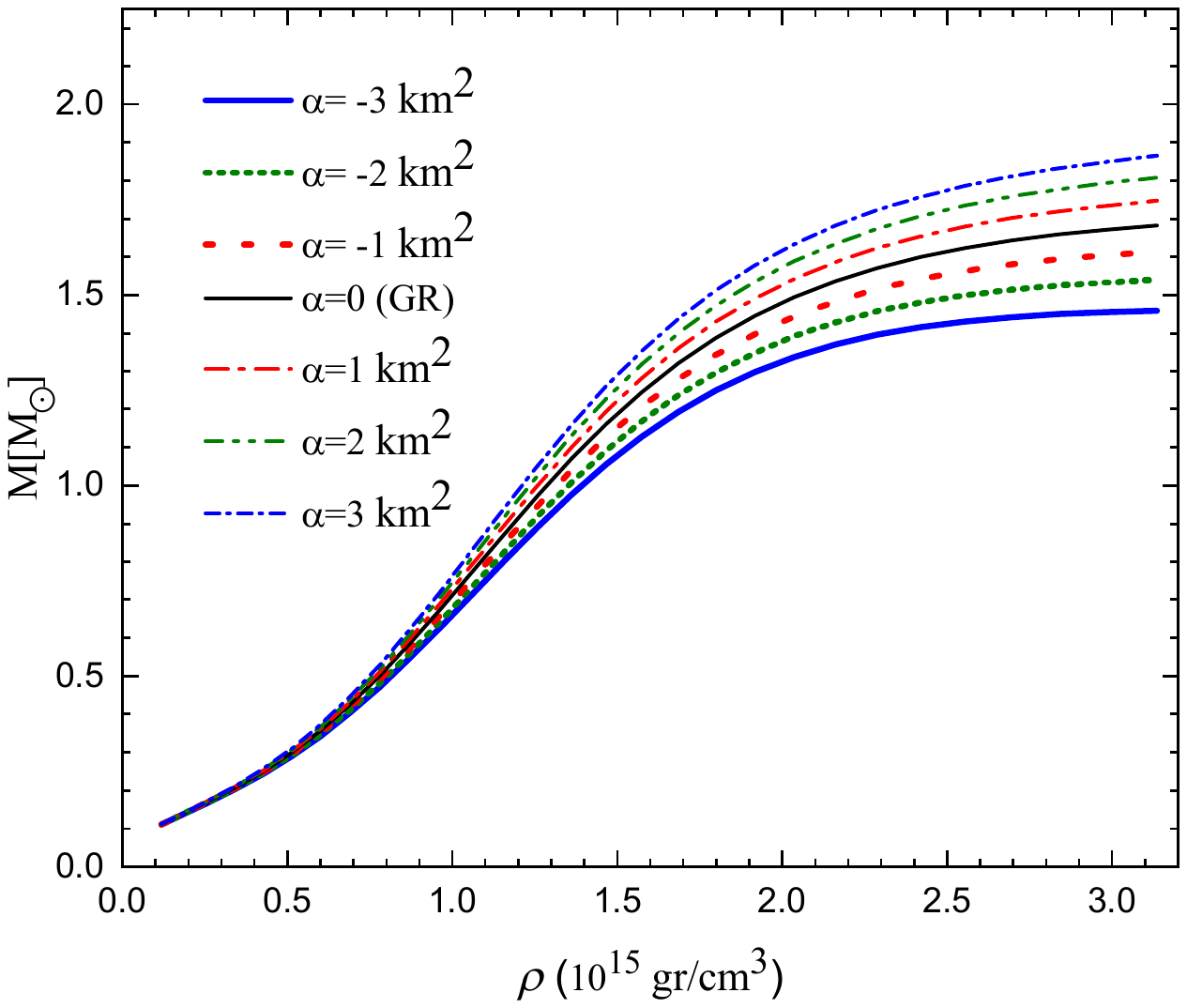}
			\includegraphics[width=9.5cm]{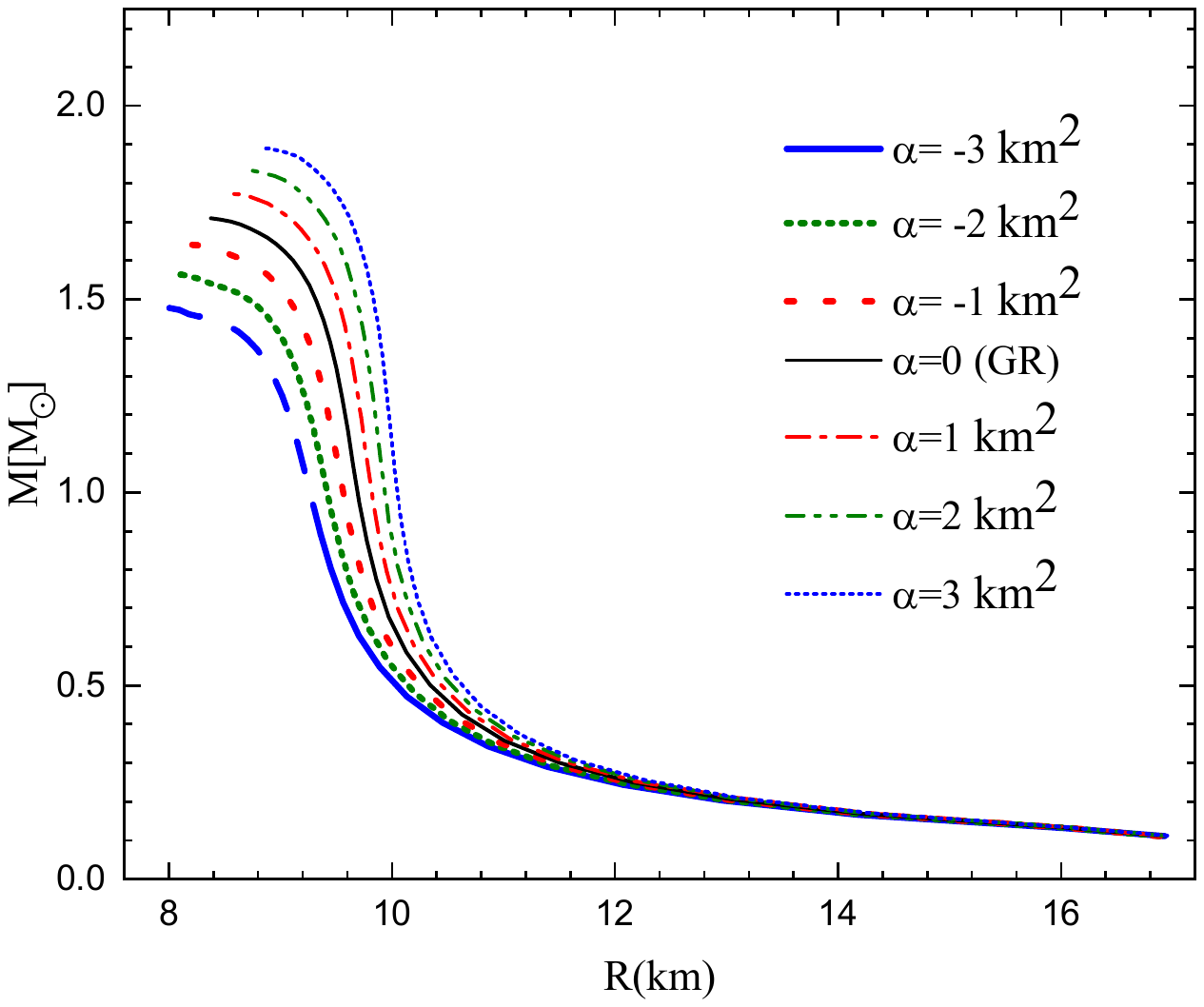}}
		\caption{{\protect\small M-$\rho$ relation (up panel) and M-R  relation (down panel)  of isotropic non-magnetized neutron star for different coupling constants ($\alpha$) in 4D EGB gravity.}}
\label{alpha0}
\end{center}
\end{figure}
%
%
%

\subsection{RESULTS AND DISCUSSIONS}
By solving the stellar structure equations for a range of central density values, we can classify a family of neutron stars (NSs) in 4D EGB gravity. Fig. \ref{alpha0}  illustrates the results of this study. It is evident that the coupling constant affects the maximum mass of isotropic non-magnetized neutron stars. Consequently, as the coupling constant increases, so does the gravitational mass and the corresponding radius.
For $\alpha=0$, the  4D EGB corresponding graphs agree with Einstein gravity. In addition, the 4D EGB gravity diagrams satisfy the requirement for neutron star stability, i.e., the positiveness of the derivative of the gravitational mass function with respect to the central density.
The maximum mass and corresponding radius values of neutron stars associated with 4D EGB gravity tend to approach each other as the coupling constant increases.
\begin{table*}
\caption{Properties of isotropic non-magnetized neutron stars for different coupling constants ($\alpha$) in 4D EGB gravity.}
\label{4proalpha}
\begin{tabular*}{\textwidth}{@{\extracolsep{\fill}}lrrrrrl@{}}
\hline
$\alpha\ (km^{2})$ & \multicolumn{1}{c}{ ${M_{max}}\ (M_{\odot})$} & \multicolumn{1}{c}{$R\ (km)$} & \multicolumn{1}{c}{$R_{Sch}\ (km)$} & \multicolumn{1}{c}{ $\sigma (\frac{R_{Sch}}{R})$} & \multicolumn{1}{c}{$z$} & \multicolumn{1}{c}{$K(10^{-8}$ $m^{-2})$} \\
\hline
$-3.0$ & $1.46$ & $8.18$ & $4.71$ & $0.535$ & $0.575$ & $2.97$ \\
$-2.0$ & $1.54$ & $8.38$ & $4.97$ & $0.592$ & $0.567$ & $2.92$ \\
$-1.0$ & $1.61$ & $8.56$ & $5.19$ & $0.606$ & $0.594$ & $2.87$ \\
$0.0$ & $1.68$ & $8.73$ & $5.43$ & $0.621$ & $0.625$ & $2.82$ \\
$+1.0$ & $1.75$ & $8.89$ & $5.64$ & $0.635$ & $0.655$ & $2.78$ \\
$+2.0$ & $1.81$ & $9.04$ & $5.84$ & $0.646$ & $0.680$ & $2.74$ \\
$+3.0$ & $1.86$ & $9.18$ & $5.99$ & $0.653$ & $0.699$ & $2.68$ \\
\hline
\end{tabular*}
\end{table*}

In the next step, we examine how the coupling constant affects the maximum mass, radius, and other structural properties of the neutron star. In $D=4$, our results are then compared with the structure of a neutron star without the coupling constant. The results of this study are presented in Table \ref{4proalpha}. According to the values in this table, the maximum mass and the corresponding radius increase as the coupling constant increases in the 4D EGB gravity. In other words, higher-order curvature terms (here the second order GB term) lead to more massive stars.

The next step is to examine how the coupling constant affects other structural properties of neutron stars. In the selected spacetime Eq. (\ref{metric}), structural properties such as the modified Schwarzschild radius, compactness, and surface redshift  are defined as follows,
\begin{eqnarray}
R_{Sch}=\frac{GM}{c^{2}}+\sqrt{\frac{G^2M^2}{c^{4}}-\alpha G},
\end{eqnarray}
\begin{eqnarray}
\sigma =\frac{R_{Sch}}{R},
\end{eqnarray}
\begin{eqnarray}
z&=&\frac{1}{\sqrt{g_{00}(R)}}-1\nonumber\\&=&e^{-\Phi(R)}-1\nonumber\\&=&\frac{1}{\sqrt{1-\frac{R^2}{2 \alpha G}\bigg(\sqrt{1+\frac{8\alpha G^2 M}{c^2 R^3}}-1\bigg)}}-1.
\end{eqnarray}
%
The results of these properties are presented in Table \ref{4proalpha}. According to the results, any increase in the coupling constant leads to a greater Schwarzschild radius, compactness, and surface redshift. There is no black hole in the system under investigation, because the corresponding radius is greater than the Schwarzschild radius, proving that it is definitely a neutron star.

The next property to be examined is the Kretschmann scalar. It is somewhat more complicated to analyze the Kretschmann scalar in 4D EGB gravity, because it differs from Einstein's gravity. Initially, we have to decide which of the curvature scalars should be used to measure the neutron star curvature. As we know, the Ricci tensor and Ricci scalar components vanish outside the star. In vacuum, the two quantities are zero because of their dependance on pressure and density.
To calculate the space-time curvature, we need to use another quantity. In comparison to the Ricci scalar and components of the Ricci tensor, the non-zero components of the Riemann tensor are more suitable measures of space-time curvature. For simplicity, we can evaluate the Kretschmann scalar to measure the vacuum curvature instead of using the Riemann tensor. In Einstein gravity, the surface curvature of a neutron star is given by \cite{Eksi},
\begin{equation}
K=%
\sqrt{R_{\mu \nu \alpha \beta }R^{\mu \nu \alpha \beta }}=\frac{4\sqrt{3}GM}{%
c^{2}R^{3}}	
\end{equation}%
In 4D EGB gravity, we can calculate the Kretschmann scalar as follows:
\begin{eqnarray}
K&=&
\sqrt{R_{\mu \nu \alpha \beta }R^{\mu \nu \alpha \beta }}\nonumber\\
&=&2G \biggl(\frac{12 m(r)^{2}}{r^{6}c^{4}}-\frac{4 r^{2} m(r) m^{'}(r)}{r^{6}(c^{4} r -2 c^{2} G m(r))} \nonumber\\
&& +\frac{r^{4}m^{'}(r)^{2}}{r^{6}(c^{2} r -2 G m(r))^2}\biggl)^{1/2}.
\label{kre4ga}
\end{eqnarray}%
The value of $K$ is computed at $r = R$. In Eq. (\ref{kre4ga}), we arrive at the surface curvature relation for Einstein gravity by setting the derivative of mass to zero (equivalent to setting the coupling constant to zero).
The results of  Kretschmann scalar in 4D EGB gravity for different values of coupling constants ($\alpha$) confirm that increasing $\alpha$ leads to decreasing gravitational strength (See Table \ref{4proalpha} for more details).
\begin{table*}
\caption{Comparison of our results for the mass and radius of  isotropic neutron stars with those of observation.}
\label{Comalpha}
\begin{tabular*}{\textwidth}{@{\extracolsep{\fill}}lrrrrl@{}}
\hline
$Name$ & \multicolumn{1}{c}{${M}\ (M_{\odot})$} & \multicolumn{1}{c}{$R\ (km)$} & \multicolumn{1}{c}{$Our\ work $} & \multicolumn{1}{c}{${M}\ (M_{\odot})$} & \multicolumn{1}{c}{$R\ (km)$} \\
\hline
$PSR\ J0740+6620$ & $2.10$ & $12(\pm2)$ & $For\ \alpha=7.7\ km^{2}$ & $2.10$ & $9.76$ \\
$PSR\ J0348+0432$ & $2.01$ & $13(\pm2)$ & $For\ \alpha=5.9\ km^{2}$ & $2.01$ & $9.55$ \\
$PSR\ J1614-2230$ & $1.97$ & $12(\pm2)$ &  $For\ \alpha=5\ km^{2}$ & $1.97$ & $9.44$ \\
$Vela\ $X-1$$ & $1.80$ & $11(\pm2)$ & $For\ \alpha=2\ km^{2}$  & $1.80$ & $9.04$ \\
$4U\ 1608-52$ & $1.74$ & $9(\pm1)$ & $For\ \alpha=1\ km^{2}$ & $1.75$ & $8.89$ \\

\hline
\end{tabular*}
\end{table*}

The observational results of some objects ($4U\ 1608-52$ \cite{Guver},  $Vela\ X-1$ \cite{Rawls}, $PSR\ J1614-2230$ \cite{Demorest},  $PSR\ J0348+0432 $\cite{Antoniadis} and $PSR\ J0740+6620$ \cite{Cromartie}) have been compared with the results of our calculations in Table \ref{Comalpha} to consider their results as constraints in the calculations of the neutron star structure in 4D EGB gravity. The observational results impose some constraints on the values of coupling constant ($\alpha$) to match the results of calculations with those of the observations.
For neutron stars with masses below $1.68$, the coupling constant ($\alpha$) in 4D EGB gravity must be negative (See Table \ref{4proalpha} for more details).
In order for the neutron star radius to be between $\sim (11-13)\ km$ at mass of about $\sim 1.4M_{\odot}$, the coupling constant should be in the range $14 \lesssim \alpha \lesssim 43$.



In the next section, we study the anisotropic magnetized neutron star and calculate its maximum mass and corresponding radius using generalized anisotropic TOV equations in 4D EGB gravity.

\section{The structure of an anisotropic magnetized neutron star in 4D EGB gravity}
In this part of our study, we investigate the structure of anisotropic magnetized neutron stars using a 4D EGB gravity model. In the below part, we introduce the momentum energy tensor of anisotropic matter and present generalized anisotropic TOV equations for 4D EGB gravity.
\subsection{MODIFIED ANISOTROPIC TOV EQUATIONS IN 4D EGB
GRAVITY}
The matter source is assumed to be anisotropic, whose energy-momentum tensor is given by \cite{4Tangphati},
\begin{equation}
T_{\mu \nu}=(\epsilon+P_{t})u_{\mu}u_{\nu}+P_{t}g_{\mu \nu}-\Delta k_{\mu}k_{\nu},
\label{aniemt}
\end{equation}
where $P_{r}$ is the radial pressure, $P_{t}$ is the tangential pressure, $\Delta \equiv P_{t}-P_{r}$
is the anisotropy factor and $k^{\mu}$ is a unit spacelike $4$-vector
along the radial coordinate. Here, $k^{\mu}$ is defined as $k^{\mu}=(0,e^{-\lambda(r)},0,0)$ with $k^{\mu}k_{\mu}=+1$ and $k^{\mu}u_{\mu}=0$.

According to metric in Eq. (\ref{metric}) and the anisotropic energy-momentum tensor \ref{aniemt}, we have the following anisotropic field equations,
\begin{eqnarray}
\frac{2}{r}\frac{d\lambda}{dr}=e^{2\lambda}\frac{\frac{8\pi G \rho}{c^2}-\frac{(1-e^{-2\lambda})}{r^2}(1-\alpha \frac{(1-e^{-2\lambda})}{r^2})}{(1+2\alpha\frac{(1-e^{-2\lambda})}{r^2})},
\end{eqnarray}%
\begin{eqnarray}
\frac{2}{r}\frac{d\Phi}{dr}=e^{2\lambda}\frac{\frac{8\pi G P_{r}}{c^4}+\frac{(1-e^{-2\lambda})}{r^2}(1-\alpha \frac{(1-e^{-2\lambda})}{r^2})}{(1+2\alpha\frac{(1-e^{-2\lambda})}{r^2})}.
\end{eqnarray}%
Additionally, the non-trivial radial component ($\nu=1$) of the conservation law of the anisotropic energy-momentum tensor (continuity equation) obtained from relation $\nabla_{\mu}T^{\mu}_{\hspace{2mm}\nu}=\partial_{\mu}T^{\mu}_{\hspace{2mm}\nu}+\Gamma^{\mu}_{\rho\mu}T^{\rho}_{\hspace{2mm}\nu}-\Gamma^{\rho}_{\mu\nu}T^{\mu}_{\hspace{2mm}\rho}=0$, can be gotten by the following relation,
\begin{eqnarray}
\frac{dP_{r}}{dr}=-(\rho c^2 +P_{r} )\frac{d\Phi}{dr}+\frac{2\Delta }{r}.
\label{anisodelta}
\end{eqnarray}%
{In the case of anisotropic matter, this is the generalization of Tolman-Oppenheimer-Volkoff (TOV) equations. Eq. (\ref{anisodelta}) considers the anisotropic pressure ($\Delta$) to describe the distribution of matter and pressure within the stars. The application of hydrostatic equilibrium equation for anisotropic matter allows researchers to understand how the anisotropic nature of the fluid and the presence of a strong magnetic field influence the structural properties of compact objects such as neutron stars. }

In order to study the anisotropic magnetized neutron star, we need the generalized anisotropic TOV equations. Here, we replace $e^{-2\lambda(r)}$ in the metric function with $e^{-2\lambda(r)}=1-\frac{r^2}{2 \alpha G}\bigg(\sqrt{1+\frac{8\alpha G^2 m(r)}{c^2 r^3}}-1\bigg)$ (see Ref. \cite{Doneva}, for more details).
For 4D EGB gravity after some calculations, the generalized anisotropic TOV equations are obtained as follows, 
\begin{eqnarray}\nonumber
\frac{dP_{r}}{dr}&=&-\frac{r}{2}\frac{( \rho
(r)c^{2}+ P_{r} )}{1-\frac{r^2}{2 \alpha G}\bigg(\sqrt{1+\frac{8\alpha G^2 m(r)}{c^2 r^3}}-1\bigg)} \\&&\times\frac{\frac{8\pi GP_{r}}{c^4}+\frac{1}{2\alpha G}\bigg(\sqrt{1+\frac{8\alpha G^2 m(r)}{c^2 r^3}}-1\bigg)\biggl(1-\frac{1}{2G}\bigg(\sqrt{1+\frac{8\alpha G^2 m(r)}{c^2 r^3}}-1\bigg)\biggr)}{1+\frac{1}{G}\bigg(\sqrt{1+\frac{8\alpha G^2 m(r)}{c^2 r^3}}-1\bigg)}+\frac{2\Delta }{r},
\end{eqnarray}%
\begin{eqnarray}\nonumber
\frac{dm(r)}{dr}&=&-\frac{m(r)}{r}+\frac{c^2 r^2}{2G}\biggl(\frac{\frac{8\pi G \rho(r)}{c^2}-\frac{1}{2\alpha G}\bigg(\sqrt{1+\frac{8\alpha G^2 m(r)}{c^2 r^3}}-1\bigg)\biggl(1-\frac{1}{2G}\bigg(\sqrt{1+\frac{8\alpha G^2 m(r)}{c^2 r^3}}-1\bigg)\biggr)}{1+\frac{1}{G}\bigg(\sqrt{1+\frac{8\alpha G^2 m(r)}{c^2 r^3}}-1\bigg)}\biggr)\\&&+\frac{c^2 r^2}{2 \alpha G^2 }\bigg(\sqrt{1+\frac{8\alpha G^2 m(r)}{c^2 r^3}}-1\bigg). \label{tov2}
\end{eqnarray}%
The above equations describe the hydrostatic equilibrium in 4D EGB gravity for anisotropic magnetized neutron stars. By setting $\alpha=0$, these equations are reduced to the ordinary TOV equations for an anisotropic magnetized neutron star. In the next section, we introduce a model for the anisotropy ($\Delta$) of a magnetized neutron star to solve the anisotropic TOV equations.

\subsection{Model for anisotropy}
We need a functional form for anisotropy to close the system of equations so that we can include any effects  caused by local anisotropy of the fluid and the presence of a strong magnetic field to solve the anisotropic TOV equations in 4D EGB gravity. Despite the existence of microscopic theories, there is no explicit and straightforward method to explain the combined effects of anisotropy on fluids and magnetic fields.

\begin{figure}[h!]
\begin{center}{\includegraphics[width=9.5cm]{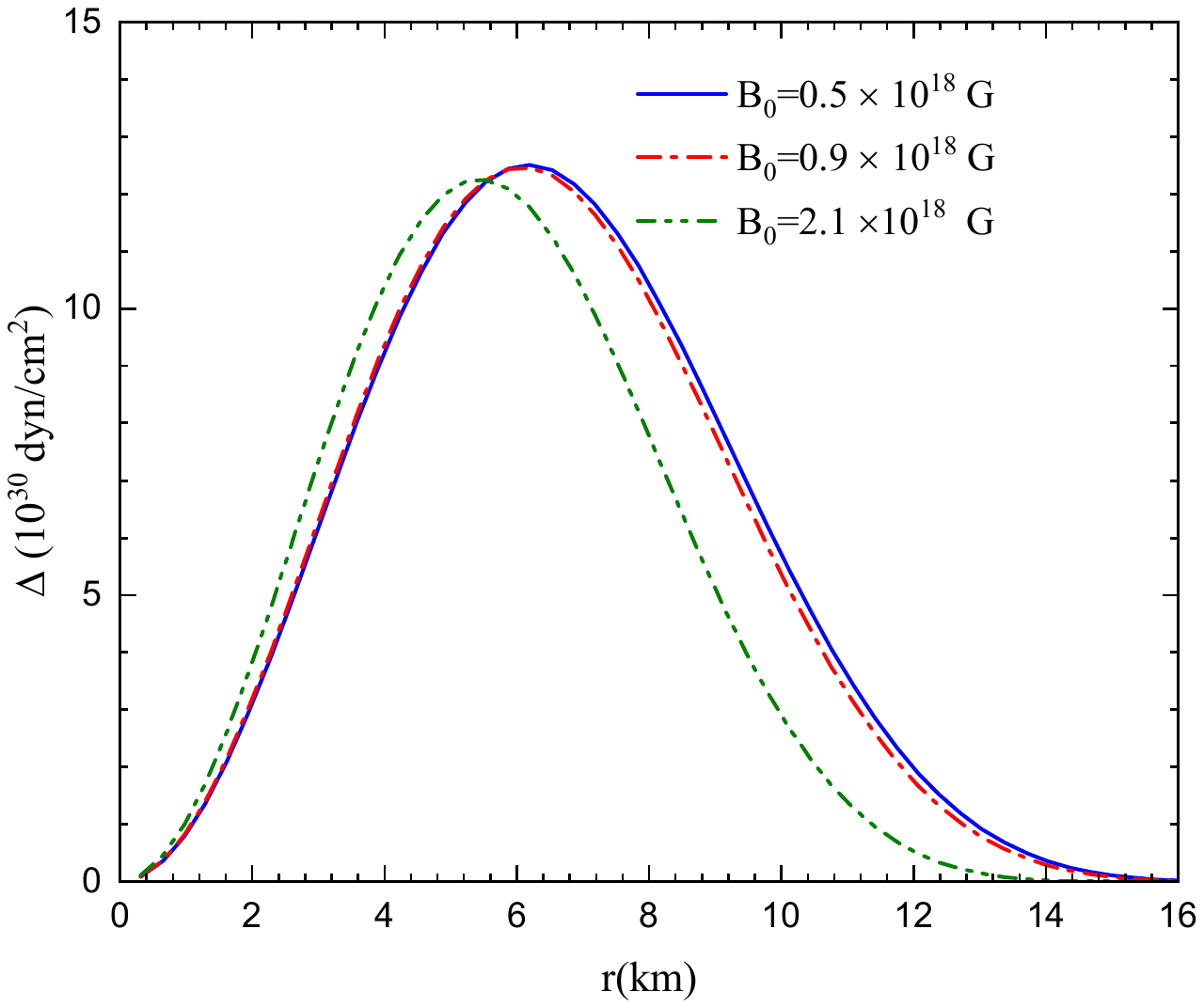}}
\caption{{\protect\small Anisotropy functions in 4D EGB gravity vs distance from the neutron star center for different $B_{0}$.}}
\label{anifacor}
\end{center}
\end{figure}

An anisotropic neutron star can be constructed using the following basic hypotheses;
a) At the center of a neutron star, the sum of hydrodynamic and gravitational forces is zero, therefore, to maintain stability through the balance of forces, anisotropy must be zero.
b) In neutron stars, anisotropy should vary with position and its dependance on radial pressure should be nonlinear.
c) The study conducted in Ref. \cite{DEB} suggests that the functional form of anisotropy includes the effects of local anisotropy of the fluid and strong magnetic fields.
Following Bower and Liang \cite{BowerLiang}, we consider the anisotropy function ($\Delta$) in the following form,

\begin{equation}
\Delta=\frac{\beta_{BL}G}{c^{4}}({c^{2}\rho+P_{r}})({c^{2}\rho+3 P_{r}})e^{2\lambda}r^{2},
\label{delta}
\end{equation}%
where $e^{-2\lambda}\approx 1-\frac{2Gm}{c^{2}r}$, and from Eq. (\ref{Pr}), we can calculate the radial pressure ($P_{r}$).
By setting $\beta_{BL}=0$, we return to the isotropic state (see Ref. \cite{BowerLiang} for more details).
In Fig. \ref{anifacor}, we plot the anisotropy factor inside the neutron star in terms of the distance from the center of the neutron star for a specific value of $\beta_{BL}$, and different values of $B_{0}$ in 4D EGB gravity.
In this figure, one can see that the anisotropy function has all the basic assumptions listed above. According to this figure, as $B_{0}$ increases, the maximum of the anisotropy function occurs at a smaller distance from the center of the star, and its maximum value decreases.
%



\subsection{Effect of anisotropy parameter on magnetized neutron star structure}

\begin{figure}[h!]
\begin{center}{\includegraphics[width=9.5cm]{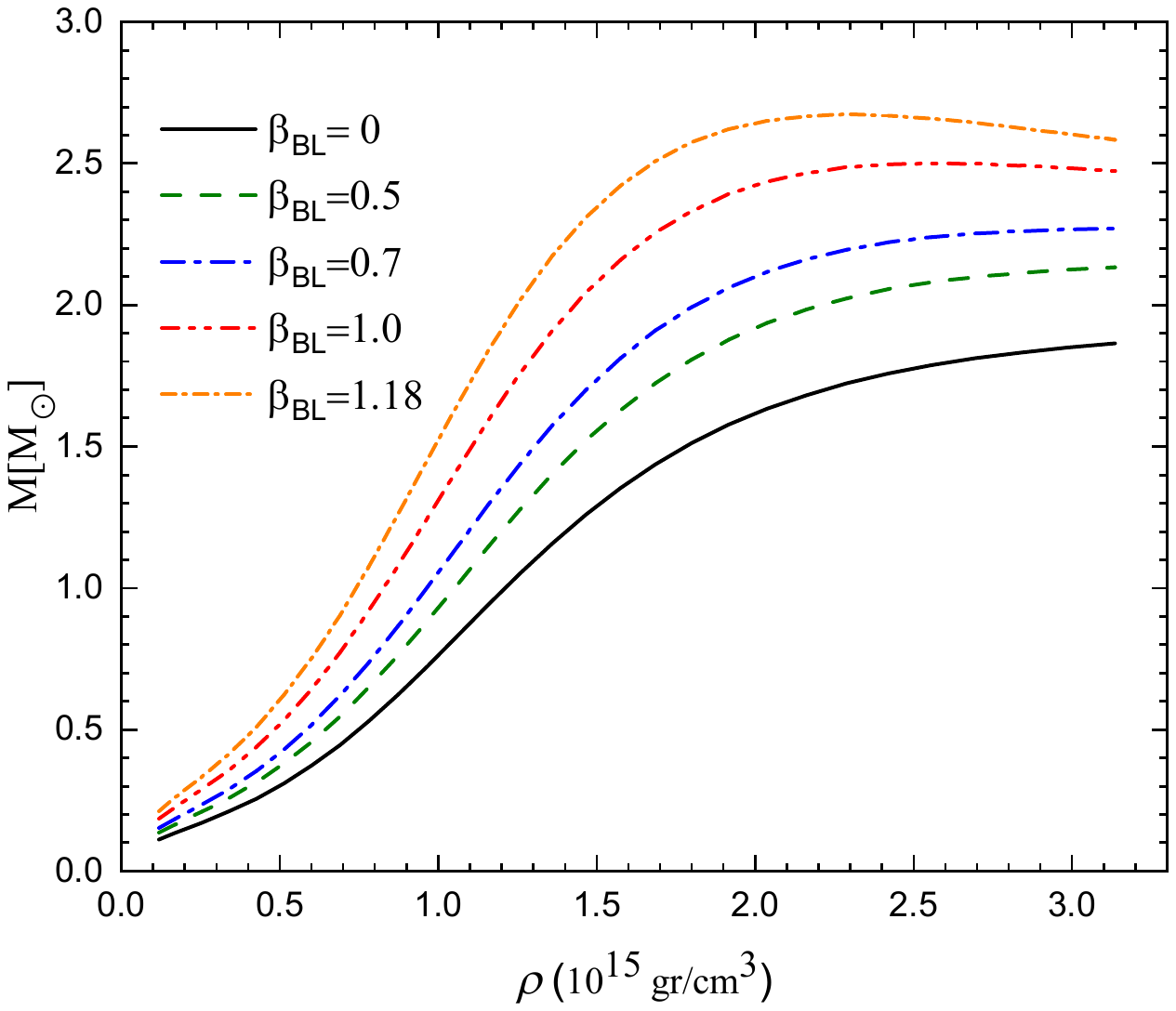}
	\includegraphics[width=9.5cm]{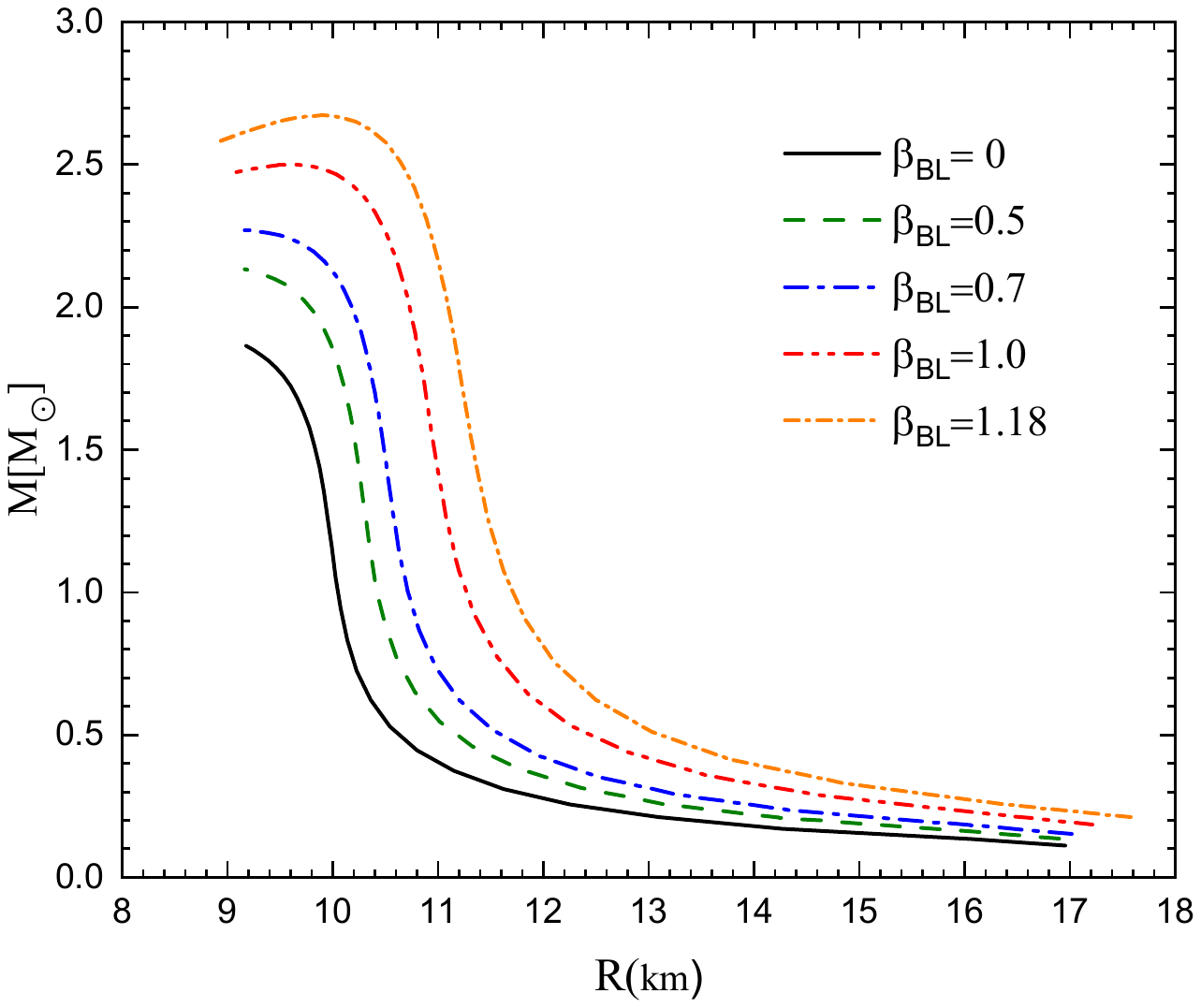}}
\caption{{\protect\small Mass-density relation (up panel) and Mass-radius  relation (down panel)  of magnetized neutron star for different values of anisotropy parameter, $B_{0}=0.9 \times 10^{18}\ G$, and $\alpha=3\ km^2$ in 4D EGB gravity. }}
\label{betadiff}
\end{center}
\end{figure}

For specific values of the coupling constant ($\alpha=3\ km^2$) and magnetic field ($B_{0}=0.9 \times 10^{18}G$), we examine the effects of the anisotropy parameter on the mass-radius and mass-density relations in 4D EGB gravity. The plots in Fig. \ref{betadiff} present the results of this investigation for different values of the anisotropy parameter $\beta_{BL}$. According to the plots in Fig. \ref{betadiff}, the anisotropy parameter significantly influences the mass-radius and mass-density curves. As a result, the maximum mass increases with increasing anisotropy.

Additionally, we examined how the anisotropy parameter affects the neutron stars maximum mass, radius, and other structural properties for $\beta_{BL}\leq 0.20$ (See Table \ref{Tbetadiff} for more details).
\begin{table*}
\caption{Properties of anisotropic magnetized neutron stars for different anisotropy parameters ($\beta_{BL}$) in 4D EGB gravity, in $B_{0}=0.9 \times 10^{18}\ G$ and $\alpha=3 km^2$.}
\label{Tbetadiff}
\begin{tabular*}{\textwidth}{@{\extracolsep{\fill}}lrrrrrl@{}}
\hline
$\beta_{BL}$ & \multicolumn{1}{c}{ ${M_{max}}\ (M_{\odot})$} & \multicolumn{1}{c}{$R\ (km)$} & \multicolumn{1}{c}{$R_{Sch}\ (km)$} & \multicolumn{1}{c}{ $\sigma (\frac{R_{Sch}}{R})$} & \multicolumn{1}{c}{$z$} & \multicolumn{1}{c}{$K(10^{-8}$ $m^{-2})$} \\
\hline
$0(with\ B=0)$ & $1.86$ & $9.18$ & $5.99$ & $0.653$ & $0.70$ & $2.68$ \\
$+0.05$ & $1.87$ & $9.08$ & $6.03$ & $0.664$ & $0.72$ & $2.79$ \\
$+0.10$ & $1.88$ & $9.09$ & $6.06$ & $0.667$ & $0.73$ & $2.80$ \\
$+0.15$ & $1.91$ & $9.11$ & $6.16$ & $0.676$ & $0.76$ & $2.82$ \\
$+0.20$ & $1.94$ & $9.12$ & $6.26$ & $0.686$ & $0.78$ & $2.86$ \\
\hline
\end{tabular*}
\end{table*}
The results of our calculations have been compared to the observational data for objects (PSR 2215+5135 \cite{Linares}, GW190814 \cite{abbott19}
  in Table \ref{comani} to consider observational values as constraints in the calculations of the anisotropic neutron star structure in 4D EGB gravity. Here we see that in order to have a good agreement between the calculation results with those of observation, the observational results are limited to certain anisotropy parameters ($\beta_{BL}$).

%
\begin{table*}
\caption{Comparison of our results for the mass of  anisotropic neutron stars in 4D EGB with those of observation in $B_{0}=0.9 \times 10^{18}\ G$ and $\alpha=3 km^2$.}
\label{comani}
\begin{tabular*}{\textwidth}{@{\extracolsep{\fill}}lrrl@{}}
\hline
$Name$ & \multicolumn{1}{c}{${M}\ (M_{\odot})$} & \multicolumn{1}{c}{$Our\ work$} & \multicolumn{1}{c}{${M}\ (M_{\odot})$} \\
\hline
$PSR\ 2215+5135$ & $2.27_{-0.15}^{+0.17}$ & $ \beta_{BL}=0.70$ & $2.27$ \\
$GW190814$ & $2.50-2.67$  & $ \beta_{BL}=1.00-1.18$ & $2.50-2.67$ \\
\hline
\end{tabular*}
\end{table*}
%
In the next section, we concern with the question of whether increasing the anisotropic neutron star internal magnetic field $B_{0}$ always increases its maximum mass and the corresponding radius.

\subsection{Effect of internal magnetic field on magnetized neutron star structure}

\begin{figure}[h!]
\begin{center}{\includegraphics[width=9.5cm]{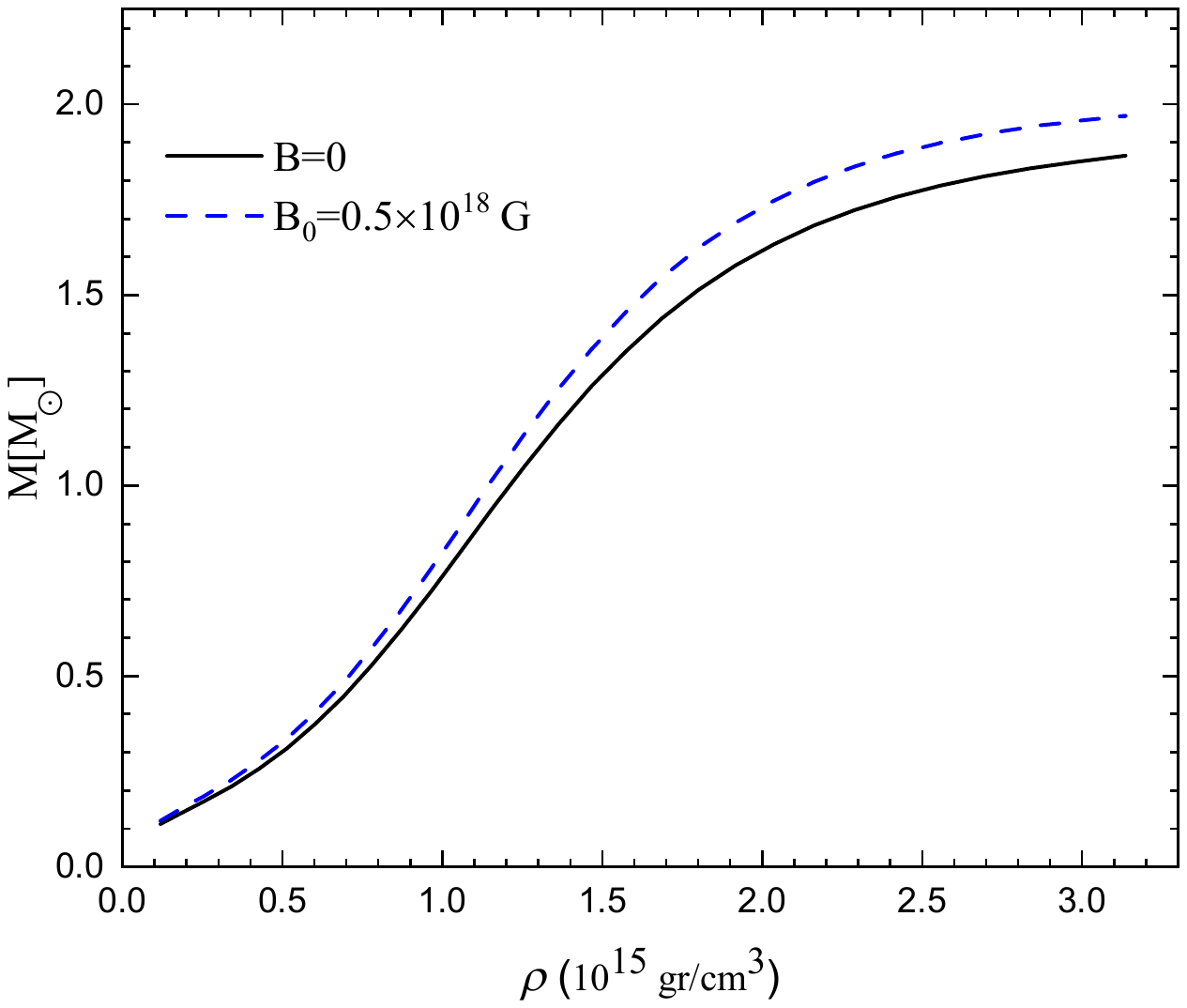}
	\includegraphics[width=9.5cm]{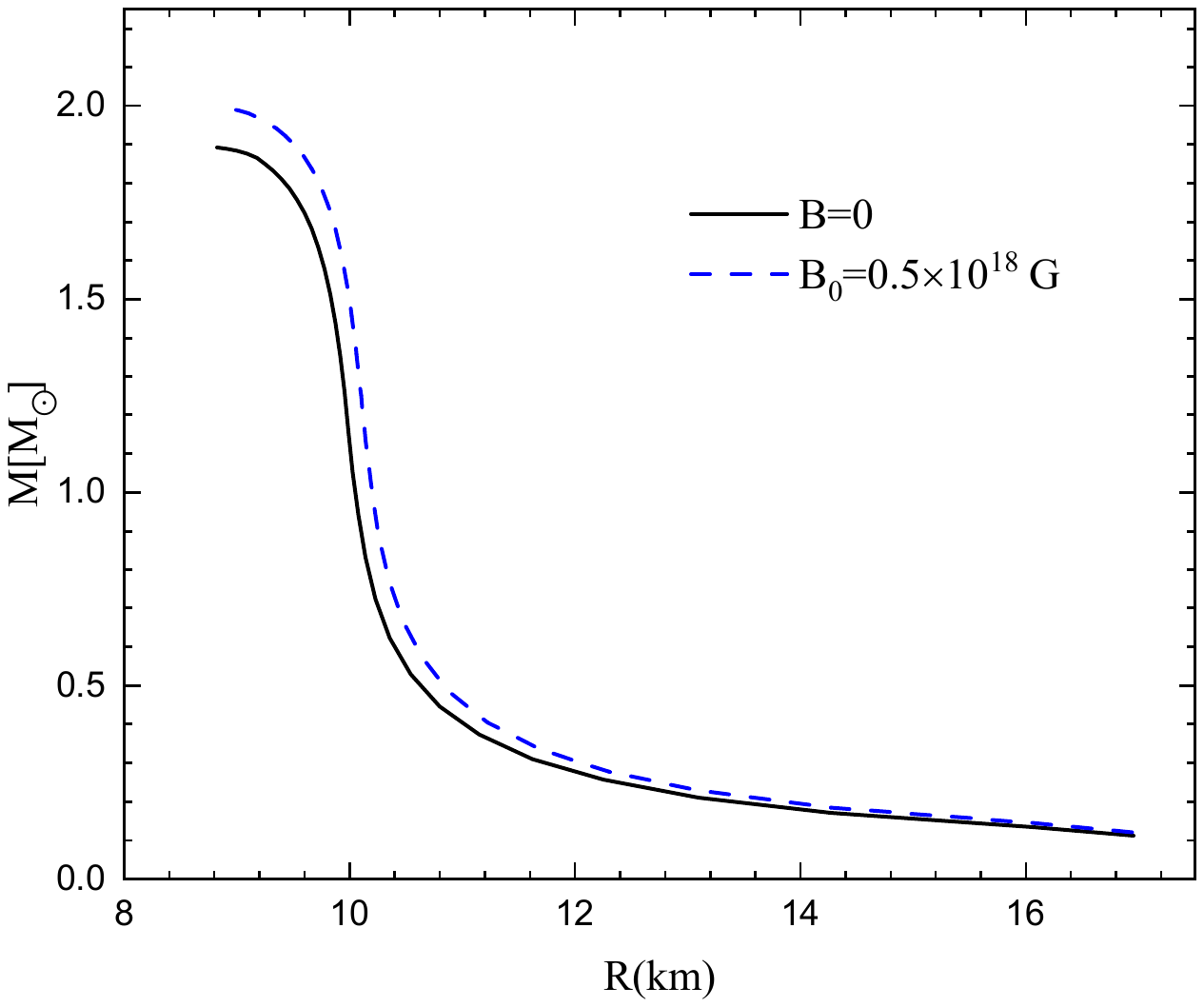}}
\caption{{\protect\small Mass-density relation (up panel) and Mass-radius  relation (down panel)  of magnetized neutron star for $B=0$ and $B_{0}=0.5\times10^{18}G$, $\alpha=3 km^2$, and $\beta_{BL}$=0.2 in 4D EGB gravity. }}
\label{MRMRO45}
\end{center}
\end{figure}

This section aims to study the effect of the internal magnetic field $B_{0}$ on the maximum mass and corresponding radius of a magnetized neutron star.
It can be seen in Fig. \ref{MRMRO45} that in the presence of the internal magnetic field $B_{0}$ of the magnetized neutron star, the maximum mass and the corresponding radius increase by increasing the magnetic field.
As shown in Table \ref{4proani}, the maximum mass, the corresponding radius and other physical properties of the magnetized neutron star are provided for various values of the internal magnetic field $B_{0}$ in 4D EGB gravity. Here, we examine the behavior of these physical properties.
%
\begin{table*}
\caption{Properties of anisotropic magnetized neutron star for different magnetic fields in 4D EGB gravity, in $\alpha=3\ km^2$ and $\beta_{BL}=0.2$.}
\label{4proani}
\begin{tabular*}{\textwidth}{@{\extracolsep{\fill}}lrrrrrl@{}}
\hline
$B_{0}(G)$ & \multicolumn{1}{c}{ ${M_{max}}\ (M_{\odot})$} & \multicolumn{1}{c}{$R\ (km)$} & \multicolumn{1}{c}{$R_{Sch}\ (km)$} & \multicolumn{1}{c}{ $\sigma (\frac{R_{Sch}}{R})$} & \multicolumn{1}{c}{$z$} & \multicolumn{1}{c}{$K(10^{-8}$ $m^{-2})$} \\
\hline
$B=0$ & $1.86$ & $9.18$ & $5.99$ & $0.653$ & $0.70$ & $2.68$ \\
$0.5\times 10^{18}$ & $1.97$ & $9.44$ & $6.35$ & $0.691$ & $0.80$ & $2.83$ \\
$0.9\times 10^{18}$ & $1.94$ & $9.12$ & $6.26$ & $0.686$ & $0.78$ & $2.86$ \\
$2.1\times 10^{18}$ & $1.76$ & $8.57$ & $5.68$ & $0.662$ & $0.72$ & $3.12$ \\
\hline
\end{tabular*}
\end{table*}
%
In our study, we found that increasing the internal magnetic field  $B_{0}$ does not always increase the maximum mass and corresponding radius of the magnetized neutron star. These tables show that for the magnetic fields $B_{0} \leqslant 0.5\times 10^{18} G$, the maximum mass and corresponding radius increase, while
for the internal magnetic fields $B_{0} \geqslant 0.5\times 10^{18} G$, the maximum mass and corresponding radius decrease.
The modified Schwarzschild radius and compactness have a behavior similar to the maximum mass and corresponding radius.
The Kretschmann scalar always increases with 4D EGB gravity as the internal magnetic fields $B_{0}$ increase.
Thus, by increasing the internal magnetic fields $B_{0}$, the strength of gravity increases.

\section{Conclusions}

This study analyzed and investigated the structure of isotropic non-magnetized and anisotropic magnetized neutron stars in 4D EGB gravity frameworks. To accomplish this, we computed the equation of state for a neutron star containing pure neutron matter  in the absence and  in the presence of a strong magnetic field. To calculate the anisotropic pressure, we used the Gaussian form of the magnetic field, which is dependent on the density of the material. Our first step was to investigate the causality of the equation of state of isotropic non-magnetized and anisotropic magnetized neutron stars. Next, we integrated the generalized TOV equations in 4D EGB gravity to determine the maximum mass of isotropic non-magnetized and anisotropic magnetized neutron stars and their corresponding radii. On the basis of these results, we calculated the Schwarzschild radius, compactness, surface redshift, and Kretschmann scalar of the isotropic non-magnetized and anisotropic magnetized neutron stars.

In the isotropic case, we found that the mass-radius, mass-energy density relation strongly depend on the coupling constant. Thus, increasing the coupling constant in 4D EGB gravity increases the maximum mass and the corresponding radius of the neutron star. In fact, the GB term (made from the integration of two orders of Ricci and Riemann tensors) creates an additional non-trivial contribution that leads to improved mass-radius and mass-energy density relations. As the coupling constant increases, the Schwarzschild radius, compactness, and surface redshift also increase. As a final result, increasing the coupling constant decreases the Krischmann scalar.

In the anisotropic case, we found that the relation of mass-radius, and mass-energy density, in addition to the coupling constant, depend on the form of the anisotropic function and the value of the internal magnetic field $B_{0}$.
An interesting observed result was that with increasing the internal magnetic field $B_{0}$, the magnetized neutron star maximum mass and the corresponding radius did not always increase.

Finally, we observed that both isotropic and anisotropic cases show a good agreement with the observational data, when physical parameters are properly selected in 4D EGB gravity.

\section*{Acknowledgments}
We wish to thank Shiraz University Research Council. We also wish to thank S. H. Hendi (Shiraz University) and B. Eslam Panah (University of Mazandaran) for their useful comments and discussions during this work.

\noindent\textbf{Data Availability Statement:} No Data associated in the manuscript



\end{document}